\documentclass[aps,prb,twocolumn,floatfix,showpacs,amssymb,superscriptaddress]{revtex4}
\usepackage[utf8]{inputenc}
\usepackage[english]{babel}
\usepackage{amsmath}
\usepackage{amsfonts}
\usepackage{amssymb}
\usepackage{graphicx}
\usepackage{float}
\usepackage{color}
\usepackage{subfigure}
\usepackage{braket}
\usepackage{epsfig}
\begin{document}

\title{Potassium-Doped Para-Terphenyl: Structure, Electrical~Transport Properties and Possible Signatures of a Superconducting Transition}

\author{N. Pinto}
\affiliation{School of Science and Technology, Physics Division, Universit\`a di Camerino, 62032 Camerino (MC), Italy}
\affiliation{Advanced Materials Metrology and Life Science Division, INRiM (Istituto Nazionale di Ricerca Metrologica), 10135 Turin, Italy}\
\author{C. Di Nicola}
\affiliation{School of Science and Technology, Chemistry Division, Universit\`a di Camerino, 62032 Camerino (MC), Italy}

\author{A. Trapananti}
\author{M. Minicucci}
\author{A. Di Cicco}
\affiliation{School of Science and Technology, Physics Division, Universit\`a di Camerino, 62032 Camerino (MC), Italy}

\author{A. Marcelli}
\affiliation{INFN - Laboratori Nazionali di Frascati, 00044 Frascati (RM), Italy}

\author{A. Bianconi}
\affiliation{RICMASS, Rome International Centre for Material Science Superstripes, 00185, Rome, Italy; antonio.bianconi@ricmass.eu}
\affiliation{Institute of Crystallography, IC, CNR, Monterotondo, 00185, Rome, Italy}

\author{F. Marchetti}
\affiliation{School of Science and Technology, Chemistry Division, Universit\`a di Camerino, 62032 Camerino (MC), Italy}

\author{C. Pettinari}
\affiliation{School of Pharmacy, Chemistry Unit, Universit\`a di Camerino, 62032 Camerino (MC), Italy}

\author{A. Perali}
\affiliation{School of Pharmacy, Physics Unit, Universit\`a di Camerino, 62032 Camerino (MC), Italy}


\date{\today}

\begin{abstract}
Preliminary evidence for the occurrence of high-$T_C$ superconductivity in alkali-doped organic materials, such as potassium-doped p-terphenyl (KPT), were recently obtained by magnetic susceptibility measurements and by the opening of a large superconducting gap as measured by ARPES and STM techniques. In this work, KPT samples have been synthesized by a chemical method and characterized by low-temperature Raman scattering and resistivity measurements. Here, we~report the occurrence of a resistivity drop of more than 4 orders of magnitude at low temperatures in KPT samples in the form of compressed powder. This fact was interpreted as a possible sign of a broad superconducting transition
taking place below $\sim$$90$ K in granular KPT. The granular nature of the KPT system appears to be also related to the $\sim$$20$ K broadening of the resistivity drop around the critical temperature.
\end{abstract}


\maketitle

\section{Introduction}
Molecules with extensive systems of $\pi$-bonds are the main components of the so-called molecular superconductors \cite{Romero2017}. 
Potentially interesting are aromatic hydrocarbon molecules, consisting of connected benzene rings, since doping by alkali metals of their solid phase provides the superconductivity \cite{Romero2017}. The first discovery of such superconductivity was made for K-doped picene~\cite{Mitsuhashi2010}. Aiming to find aromatic compounds promising for high $T_C$ superconductivity, attention was focused on $p$-terphenyl. The structure of the undoped crystalline $p$-terphenyl C$_{18}$H$_{14}$ ($p$-terphenyl in the following) has attracted long standing interest \cite{Rietveld,Baudour,Cailleau1,Lechner,Cailleau2,Baranyai,Goossens,Rice}. This molecule is made of three phenyl rings connected by single C-C bond in para position. The room temperature (RT) structure was described as a packing of nanoribbons, made of $p$-terphenyl molecules, in the crystallographic $b$~direction. Therefore, the structure of the molecule
 can be interpreted as a realization \mbox{of a nanoscale} architecture of low dimensional systems \cite{Pinto2016,Rezvani2016} described as a self-organized array of~polymeric quantum stripes \cite{Mazziotti,Bianconi1}. 
While pristine $p$-terphenyl is electrically insulating, it behaves \mbox{as a metal} upon K-doping \cite{Wang2017}. Recent experimental signatures of superconductivity in alkali-doped para-oligophenyl compounds \cite{Huang2019,Yan2019,Zhong2018a}
and in $p$-terphenyl K$_x$C$_{18}$H$_{14}$ \cite{Chen1,Chen2,Chen3,Liu1} in particular,
has attracted considerable interest. 
Increasing the K concentration toward K$_3$C$_{18}$H$_{14}$ and lowering the temperature, the K-doped $p$-terphenyl (KPT) shows signatures of superconductivity in the magnetic response, consisting \mbox{of a weak} Meissner shielding effect \cite{Chen3}. The reported range of superconducting critical temperature, $T_C$, goes from a few K \cite{Wang2017} to 123 K \cite{Chen3}, where the {repeatability} of $T_C$ is difficult since this remains a complex quantum material characterized by inhomogeneous concentrantion of dopants \mbox{and a competition} of multiple coexisting phases.
The dc and ac magnetic susceptibility measurements reported by Wang et {\it al.} demonstrate that KPT is a type-II superconductor \cite{Wang2017}. \mbox{In the KPT} compound studied in Ref. \cite{Wang2017} with a $T_C$ = 7.2 K, characterized by a moderate level of K-doping, the critical magnetic fields extrapolated at zero temperature are $H_{c1}(0)=$ 163 Oe and $H_{c2}(0)=$ 1317 Oe, from~which it results a magnetic penetration depth $\lambda=$ 76 nm and a coherence length $\xi=$ 50~nm, by~applying Ginzburg-Landau relations. In the KPT compound showing a much higher $T_C$ of the order of~120 K, with an higher level of K-doping, there is evidence for {an} upper critical magnetic field higher than 3 T, as expected for high-$T_C$ type-II superconductors \cite{Chen3}.
An independent detection of~superconductivity at 107 K in KPT compounds by magnetic measurements
was reported in~Ref.~\cite{Neha2018}.
{On the other hand, in Ref. \cite{Carrera2019} no signs of a superconducting transition have been detected by magnetic susceptibility and resistivity measurements in a KPT compound. Indeed, a~metallic behaviour was found below 150 K. However,} high resolution angle resolved photoelectron spectroscopy (ARPES) technique measurements, realized on potassium surface-doped $p$-terphenyl  demonstrate the opening of a sizeable gap in the excitation spectrum,
of the order \mbox{of $\Delta \sim$ 12 meV}, persisting up to \mbox{120 K \cite{Dessau}}. The~gap is~almost independent on \mbox{temperature in the range} \mbox{10$\div$60 K}. Above $T=60$ K the gap fills in: the scattering rate becomes of the same order of the energy gap. \mbox{At $T$ = 120 K} the photoemission
spectrum resembles the one of a normal metal. 
The existence \mbox{of a gapped} spectrum, with an energy gap of 11 meV close to that detected by ARPES, was confirmed by scanning tunneling microscopy (STM) applied to monolayer KPT \cite{Ren}. The gap closes around $T$ = 50 K, even~though no response \mbox{to a magnetic} field up to 11 T was observed. 
\mbox{ARPES and STM} experiments point toward a superconducting gap and coherent Cooper pairing below $T$ = 60 K, while in the range 60$\div$120~K Cooper pairing could be~local and incoherent: the pseudogap physics accompanied with strong fluctuations could emerge \mbox{in a similar }way to underdoped cuprate superconductors and ultracold fermions in the crossover regime of the BCS-BEC crossover \cite{Perali1,Palestini2012,Perali2,Marsiglio2015}.
Multiband effects combined with the BCS-BEC crossover could also contribute in determining the optimal
parameter configuration for high-$T_C$ superconductivity in this material \cite{Salasnich2019,Tajima2019}.
A possible theoretical scenario of the high-$T_C$ in these systems is based on resonance effects in a superlattice of~quantum stripes associated with the quasi-1D structure of the $p$-terphenyl with coexisting polarons and Fermi particles \cite{Mazziotti,Valletta1997}.
The vibrational properties and the geometrical features of the lattice should play a key role in this system in the formation of Cooper pairs induced by strong electron-phonon coupling with high energy phonons.
The role of strong electronic correlation and of the multiband electronic structure in determing the superconducting
properties of KPT {has} been discussed in Ref.~\cite{Fabrizio2017}.\\
In this article we report the experimental investigation of the conducting and {possible} superconducting behaviour of K-doped $p$-terphenyl samples, obtained by our independent synthesis. The structure of the material was studied by Raman spectroscopy as a function of the temperature and correlated with the electrical behaviour. The huge drop of the resistivity below $\sim$$90$ K, occurring in about 20 K, suggests a granular-like superconducting behaviour.
 
\section{Sample synthesis and measurement techniques}
$p$-terphenyl and potassium metal (99$\%$ purity) have been purchased from Sigma-Aldrich.

The $p$-terphenyl was purified by sublimation, to achieve a purity above 99.9$\%$. The potassium was purified by distillation.
Using a glove box, potassium 3 mol, (117 mg) was cut in little pieces and mixed with $p$-terphenyl 1 mol, (230 mg). The mixture was sealed in a glass tube under argon atmosphere and heated at 503 K for 2 h. An oxygen and water sensitive black powder \mbox{was obtained}.

\begin{figure}[H]
\centering
\includegraphics[angle=0,width=1\columnwidth]{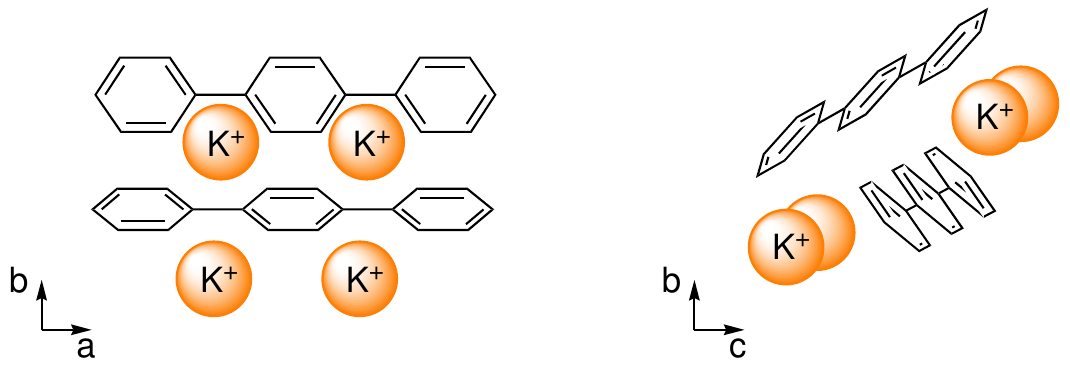}
\caption{Example of doping with two K atoms of a $p$-terphenyl molecule, K$_{2}$C$_{18}$H$_{14}$. The tilted geometry results from the first principle calculations of Ref.~\cite{Zhong2018}.} 
\label{figKPT}
\end{figure}

The exposure to atmosphere of doped $p$-terphenyl gives rise to its decomposition into a whitish hygroscopic mixture of $p$-terphenyl and potassium hydroxide, mainly due to the presence of moisture in air. As well known in the literature \cite{Connelly1996}, similar potassium-doped aromatic compounds react immediately with water affording an aqueous solution of KOH and the undoped aromatic compound with evolution of gaseous hydrogen. For this reason we have decided to take advantage of reactivity by analyzing the amount of potassium ion {which is} present in the final solution obtained by reaction of our doped sample with water, using atomic absorption spectroscopy (AAS), by a Perkin Elmer mod. 1100 B, and then go back to the molar ratio between potassium and $p$-terphenyl in the starting doped material. 
7.2 mg (0.0239 mmol) of doped $p$-terphenyl was suspended in 50 ml of distilled water and left to react at room temperature under stirring for 20 min, the
suspension was then filtered off~and analysed. The analysis has revealed 34.4 ppm of present in the final solution, corresponding to~1.72 mg of potassium in 7.2 mg of doped $p$-terphenyl sample, so the ratio (potassium):($p$-terphenyl) in milligrams is (1.72):(5.48) corresponding to a millimolar ratio (1.85):(1.00), in agreement with the presence of about 2 potassium cations for each $p$-terphenyl molecule, as shown
in Figure \ref{figKPT}.

Raman spectra were measured using a micro-Raman setup including a Czerny-Turner spectrometer (iHR320 Horiba Scientific) equipped with 1800 grooves/mm grating and an open space microscope to accommodate the cell for low temperature measurements (Linkam THMS600). Experiments were performed with an excitation wavelength of 532 nm. The laser power was set at less than 1 mW to avoid sample damage.
To measure the KPT spectra, the synthesized powder was loaded into an especially designed sample container sealed under Ar atmosphere to prevent any exposure to air and moisture.

Electrical characterization was carried out on compressed KPT powder in the range \mbox{75 $\div$ 115 MPa}. The powder was inserted into a teflon block, with a cylindrical hollow \mbox{(with a diameter} of 8 mm or 12.8 mm), and compressed by two air-tight copper leads, \mbox{part of a mechanical} vise suitably designed for this purpose, resulting in a KPT flat disk inside the teflon block. All~electrical measurements have been done keeping the KPT disk inside the mechanical vise. To each copper lead, working as electrode in tight contact with the KPT disk base, two couples of electrical cables implement the 4-wires geometry of measure. {The minimum resistance value achievable by our measuring system was checked to be $\simeq 9\times 10^{-5}$ $\Omega$, about 1.5 orders of magnitude lower than the~lowest resistance detected in the KPT material.}
Sample cooling was achieved by two different equipments, \mbox{using for each} a specific mechanical vise (as that described above) keeping disks and electrodes under a tight electrical contact during the measurement. Mechanical vises turn out to be also necessary to keep good electrical contacts among powder grains.
In one of the two equipments, the mechanical vice was located inside a small open air dewar and liquid nitrogen (LN$_2$) slowly poured.
This system was used to measure the~resistivity, $\rho(T)$, during the~cooling from 300 K to~about 80 K, occurred~in about 15~min. 
{This~apparatus was preferred for~resistivity measurement, not requiring a temperature stabilization, since it has allowed to keep as~short as possible the time elapsed from the fabrication of KPT disks to the beginning \mbox{of the electrical }characterization, generally less than one~hour.}~The~{sample temperature was measured} \mbox{by a thermocouple} and monitored \mbox{by an Eurotherm} mod. 3216 temperature controller. In~the second equipment, a He closed cycle cryostat has allowed to investigate the current-voltage characteristics at several fixed temperatures \mbox{(for details see} \cite{Pinto2018,Rezvani2019}). Measurements have been started upon cooling at {$\simeq 3$ K.}
For K doped $p$-terphenyl disks, the resistivity and the current-voltage characteristics (I-V) have been carried out sourcing a dc current by a Keithley mod. 220 and measuring the voltage drop with a Keithley mod. 2000.
Conversely, for undoped $p$-terphenyl disks, due~\mbox{to the high} resistance of the material, both resistivity and I-V characteristics have been measured in~the two contacts geometry, by using a pico-ammeter Keithley mod. 6487, applying a~constant voltage and measuring the current.\\

\section{Results and Discussion}
\subsection{Raman Spectroscopy}
\begin{figure}[H]
\centering
\includegraphics[angle=0,width=1\columnwidth]{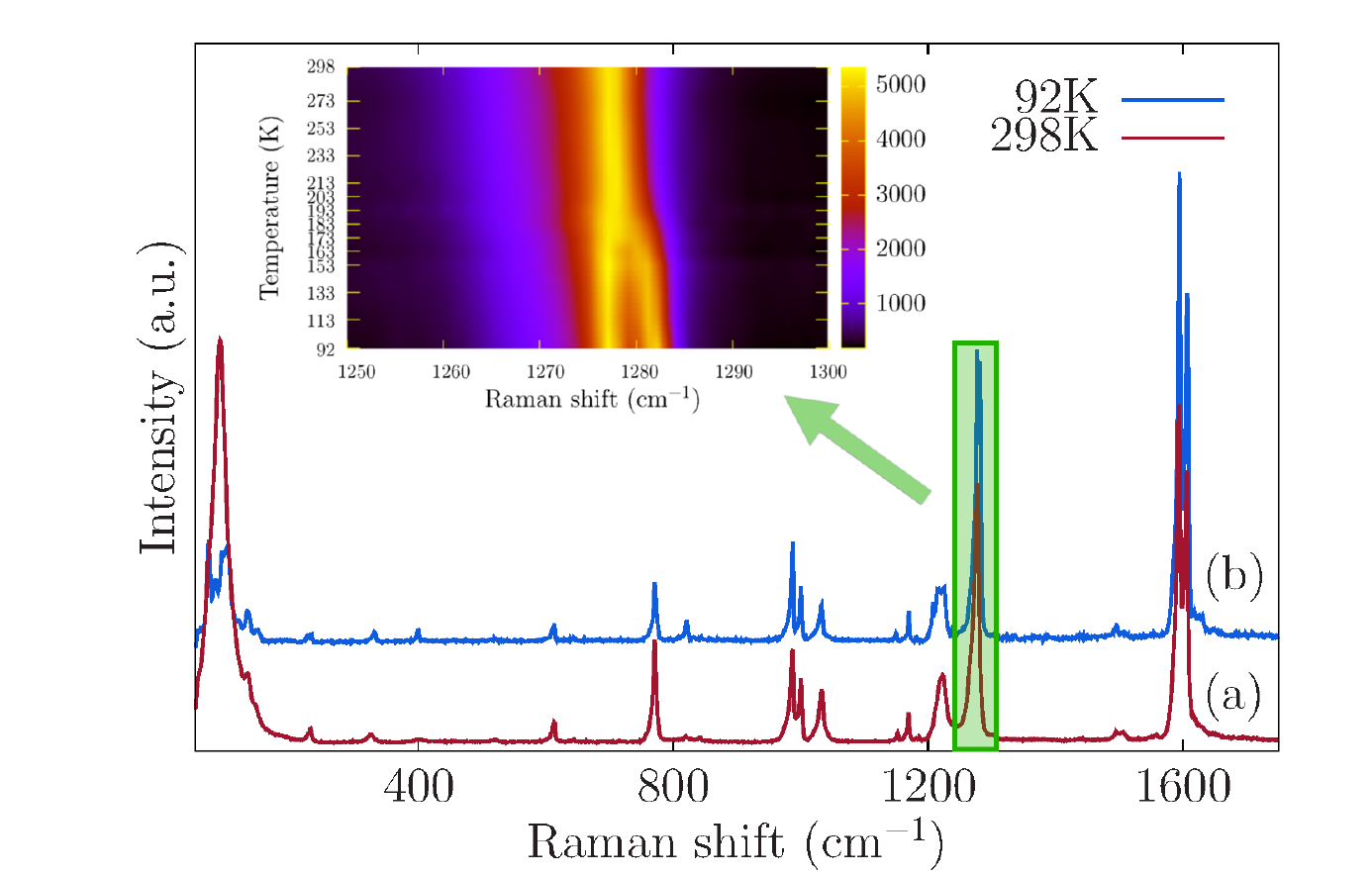}
\caption{Raman spectra of pristine $p$-terphenyl at room temperature (\textbf{a}) and 92 K (\textbf{b}). Curves have been shifted for clarity. The inset shows the evolution of the Raman band at $\sim$1280 cm$^{-1}$ (C-C inter-ring stretching) as lowering the temperature. The right scale (colour code) represents the intensity.}
\label{figRm2}
\end{figure}

\begin{figure}[H]
\centering
\includegraphics[angle=0,width=1\columnwidth]{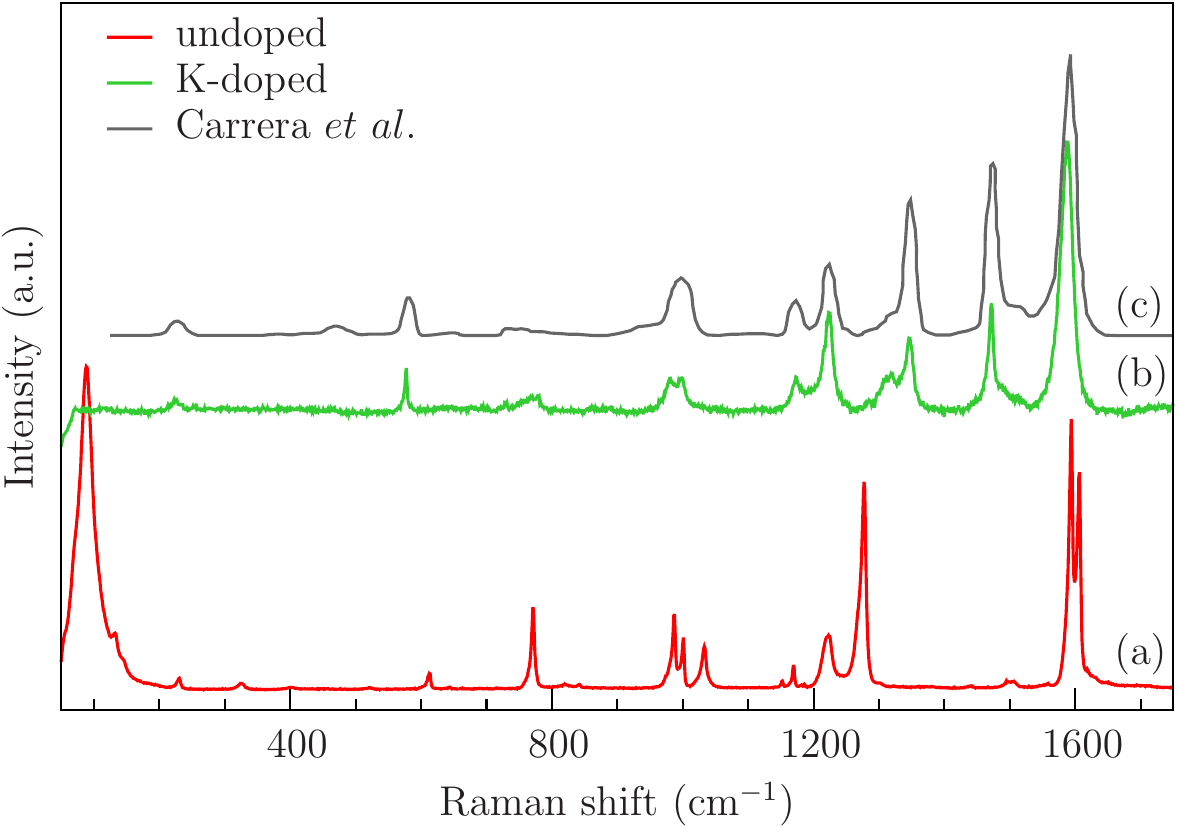}
\caption{Raman spectra of undoped (\textbf{a}) and K-doped (\textbf{b}) $p$-terphenyl measured at RT. Curves have been shifted for clarity. For comparison, the Raman spectrum for the K-doped para-terphenyl as digitized from Ref. \cite{Carrera2019} is also shown (\textbf{c}).}
\label{figRm3}
\end{figure}

\begin{figure}[H]
\centering
\includegraphics[angle=0,width=1\columnwidth]{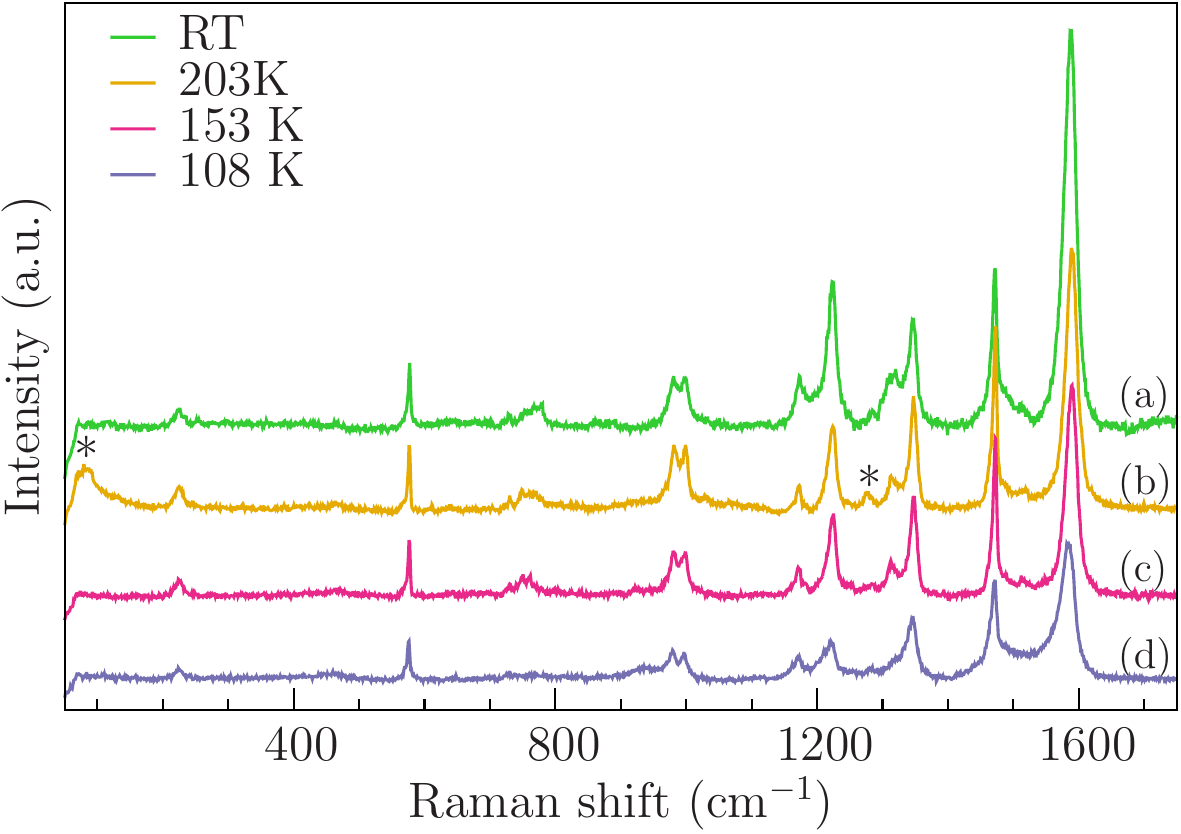}
\caption{Raman spectra for K-doped $p$-terphenyl measured at different temperatures, from RT (\textbf{a}) down to 108 K (\textbf{d}). At 203 K, two additional features, not visible in the other plotted spectra, have been marked by a star. Curves have been shifted for clarity.}
\label{figRm4}
\end{figure}

Raman spectra of undoped $p$-terphenyl have been measured in the temperature range 298 $\div$ 92 K and are shown in Figure~\ref{figRm2}. 
Fluorescence background is subtracted in all the spectra. All the peaks have been assigned to the molecular (high frequency) and lattice modes (low frequency, Raman shift $< 200$ cm$^{-1}$) \cite{Bolton1978}.
As shown in Figure~\ref{figRm2} (inset), the band at $\sim$1280 cm$^{-1}$, assigned to C-C inter-ring stretching, splits on lowering the temperature, as reported in previous studies~\cite{Amorim_da_Costa1997,Zhang2018}.
A similar behaviour is observed for the band at $\sim$230 cm$^{-1}$ associated with C-C bending modes.
Such changes are related to the disorder-order transition occurring in the $p$-terphenyl lattice at about 190 K from a~monoclinic structure to a triclinic structure \cite{Baudour1976, Baudour, Rietveld}.
At room temperature, each $p$-terphenyl molecule has a conformation planar on average, with thermal vibrations of the phenyl rings due to competing steric repulsion between H atoms (favouring a ring tilt) and delocalized $\pi$ electrons (promoting the planar geometry). Decreasing the temperature, each molecule changes into a configuration in~which the external phenyl rings are tilted and such molecular re-arrangement results in a different lattice~symmetry.\\
$p$-terphenyl doping by potassium (electron donor), induces important differences in the Raman spectrum with respect to the pristine compound (Figure~\ref{figRm3}).
The low frequency lattice modes disappear, possibly indicating reduced crystallinity of the doped sample.
In the high frequency range,  between 1100 cm$^{-1}$ and 1600 cm$^{-1}$, shifts of existing peaks and additional bands can be seen, related to strong modifications of the molecular geometry and to charged defects induced by doping.
The two peaks at~1593 cm$^{-1}$ and 1605 cm$^{-1}$ in the $p$-terphenyl (intra-ring C-C stretching) tend to merge into one only, downshifted at 1588 cm$^{-1}$. 
The peak at 1277 cm$^{-1}$ in the pristine corresponds to the one at 1348 cm$^{-1}$ in the doped compound. The upshift of such band assigned to inter-ring C-C stretching, indicates that the length of the C-C bond decreases.
Other observed bands originate from Raman-inactive modes of~the neutral $p$-terphenyl.
These effects are all related to the structural changes occurring upon doping: the molecular structure changes from benzenoid to quinoid and the phenyl rings are almost coplanar with nearly zero tilting between adjacent phenyl rings \cite{Bredas1984, Furukawa1996}.
The main Raman bands are in very good agreement with those reported in Ref. \cite{Carrera2019,Peres2005} \mbox{as well as} with the one reported in Ref. \cite{Wang2017}, providing evidence of a successful doping of our KPT sample. However, Raman spectroscopy does not allow quantifying the level of K-doping of the $p$-terphenyl molecule, known to affect the superconducting properties of KPT \cite{Mazziotti}. In fact, for a KPT material synthesised with different K to $p$-terphenyl mole ratio, but showing very similar Raman spectra, the~group of Ref. \cite{Mazziotti} has reported $T_c = 7.2$ K for a starting mole ratio of 2:1; $T_c = 43$ K and $T_c = 123$ K in Ref.s \cite {Zhong2018a} and \cite {Chen1}, respectively, for an higher mole ratio of 3:1.
In previous literature on doped conducting polyphenyls \cite{Furukawa1993, Peres2005, Dubois2003}, Raman spectroscopy was extensively used to characterize structural changes and to selectively identify bands corresponding to~polarons and bipolarons by changing the excitation wavelength from 514.5 nm to 1064 nm \cite{Furukawa1996}.
The~formation of bipolarons in KPT compounds is considered to be the driving force inducing the structural transition from benzenoid to quinoid associated with the deformation of the internal benzene ring \cite{Wang2017}. The Raman band at 1473 cm$^{-1}$ in our KPT (see our Figures \ref{figRm3} and \ref{figRm4}) was considered to be the fingerprint for the formation of bipolarons. Direct Bose-Einstein condensation of bipolarons at~high densities was proposed as a possible mechanism for high-$T_C$ superconductivity in KPT compounds~\cite{Chen3,Wang2017}. Original proposals of high-$T_C$ superconductivity driven by bipolarons have been reported in Refs. \cite{Chak1981,Alexandrov1981}.
To correlate structural changes and formation of~localized charged states, with the resistivity measurements at low temperature, Raman spectra of KPT have been measured also at different temperatures down to 108 K.
The spectra (Figure~\ref{figRm4}) do not evidence any change of the Raman activity, indicating that the doped material maintains the molecular structure upon cooling and new polaronic or bipolaronic bands are not detected at low temperatures. The~inter-ring C-C stretching band, close to 1348 cm$^{-1}$, does not split lowering the temperature, in~contrast to what was observed for the undoped compound possibly indicating an already planar configuration for the doped sample retained for the whole temperature range.
The~dopants decrease the torsional mobility of the central ring as a result of the reduced length \mbox{of the C-C} bonds between the rings.
Just the spectrum at $T$ = 203 K in Figure \ref{figRm4} shows weak lattice peaks and one peak at~$\sim$1280 cm$^{-1}$  (marked with a star) which have been attributed to a small fraction of undoped sample, intercepted by the microbeam ($\sim$2$\times$2 $\mu$m$^2$) used for micro-Raman measurements.
This indicates that the powder is not completely homogenous and there are grains of micrometric size of undoped $p$-terphenyl in our sample.

\subsection{Electrical Characterization}
The electrical properties of polycrystalline films of undoped $p$-terphenyl, a few micrometres thick,
have been investigated so far by several authors \cite{Lipinski,Staryga,Tkaczyk,Bizzarri}.
Depending on the temperature range and/or~the intensity of the applied electric field,
several transport mechanisms have been reported such as, field emission, Poole-Frenkel, variable range hopping, etc.
\cite{Lipinski,Staryga,Tkaczyk,Bizzarri}.
In a $p$-terphenyl film, both~metallic and insulating {behaviours} have been detected, above and below $\sim$ 80 K,
respectively~\cite{Tkaczyk} while, in the range from 110 K to about RT, Lipinski et {\it al.} have established that hopping
\mbox{is the main} conduction mechanism \cite{Lipinski}. These features are consistent with a high density of~traps or donor-acceptor-like sites in the band-gap of the material, causing a charge carrier transport characterized by~jumps among localized states
\cite{Lipinski,Staryga,Tkaczyk}. 
We have investigated the electrical properties of either undoped or~K-doped $p$-terphenyl disks.
In undoped $p$-terphenyl, resistivity values confirmed the insulating nature of the pristine material with a $\rho_{RT} > 10^{16}$ $\Omega$cm 
that, in the sensitivity limit \mbox{of the whole} measuring apparatus (instruments, cryogenic cables, etc.), remained practically constant also \mbox{at lower $T$}.

Upon doping, KPT samples with a RT resistivity greater than $\sim$$10^4$ $\Omega$cm do not show a metallic behaviour at low temperature indicating an unsuccessful doping process. On the contrary, when~the RT resistivity is below $\sim10^2$ $\Omega$cm a completely different $\rho_{RT}(T)$ behaviour is detected at low temperature.

\begin{figure}[H]
\centering
\includegraphics[angle=0,width=1.1\columnwidth]{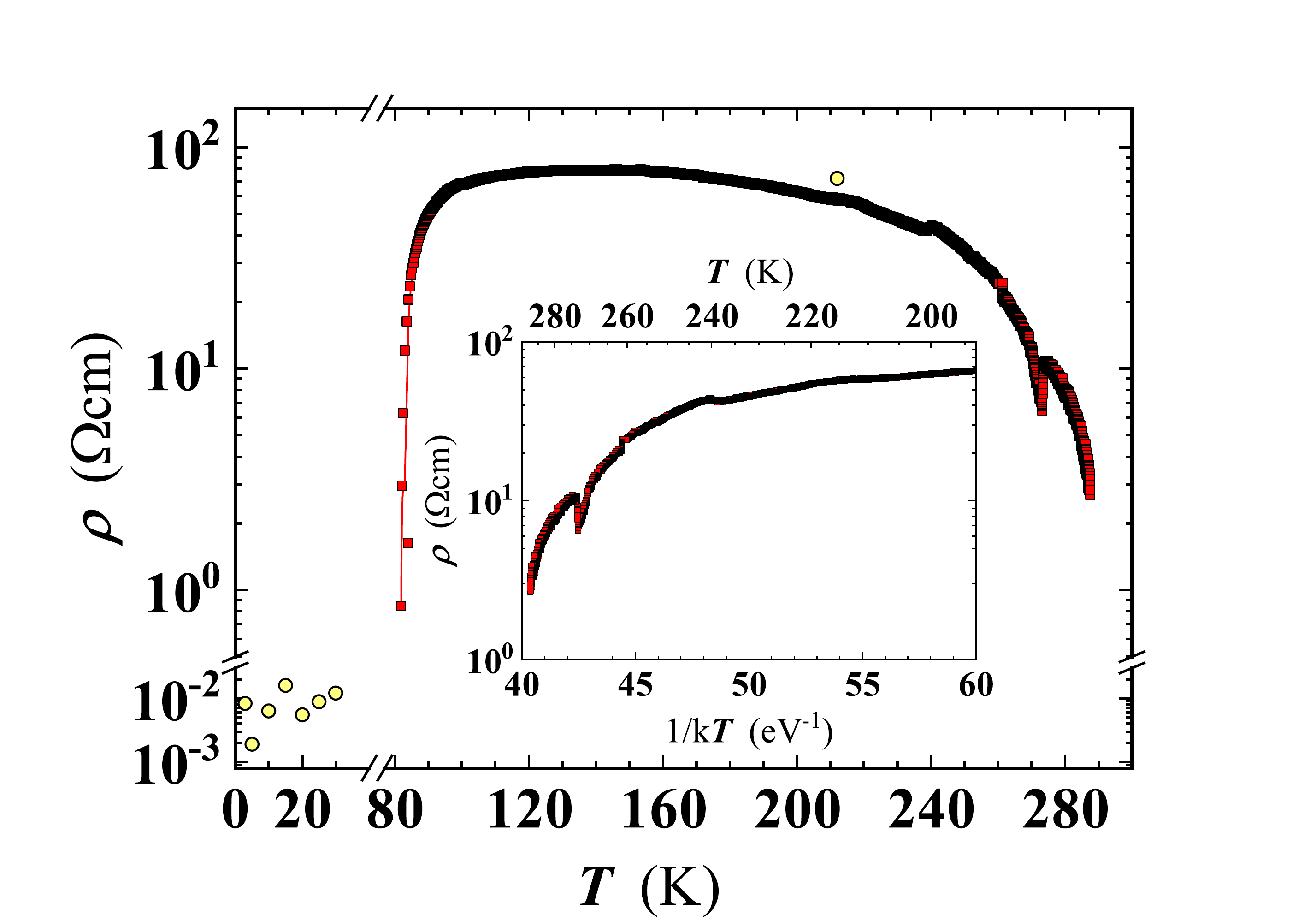}
\caption{Resistivity as a function of temperature for a doped KPT disk. {Squares: $\rho(T)$ values measured carrying out the measurement at constant current of 1$\mu$A. Circles: $\rho(T)$ values derived from I-V characteristics (see Figure \ref{fig_I-V_T}) under similar conditions of measure. Inset: Arrhenius plot of the high temperature branch of the resistivity curve.}}
\label{fig_rho_2}
\end{figure}

In detail, for {the class of} KPT samples with low $\rho_{RT}$, lowering $T$ from RT, resistivity~rises something more than one order of magnitude, saturating to $\simeq$ 80 $\Omega$cm at 130$\div$150 K. Spurious~irregularities, as peaks and bumps, are visible along this branch of the curve, \mbox{originated by a rapid} change of the pressure exerted by the mechanical vice on the KPT disk.
The~observed features in $\rho(T)$ are related to the grainy nature of KPT powder used to fabricate disks. 
Below $\approx$ 130 K, $\rho(T)$ becomes metallic-like indicating an enhanced contribution to the conduction by~electrons. The initial slow decrease of $\rho(T)$ is  followed by a steeper fall at $\sim$$90$ K that, in addition to the residual value {of less than $10^{-2}$ $\Omega$cm, below 30 K} (Figure \ref{fig_rho_2}), recalls that of a superconducting transition.
In a range of $\simeq 22$ K, a drop of resistivity by $\approx 4$ orders of magnitude was observed in our measurement. This resistivity drop shares some features of  granular superconductivity, as~discussed later.

\begin{figure}[H]
\centering
\includegraphics[angle=0,width=1\columnwidth]{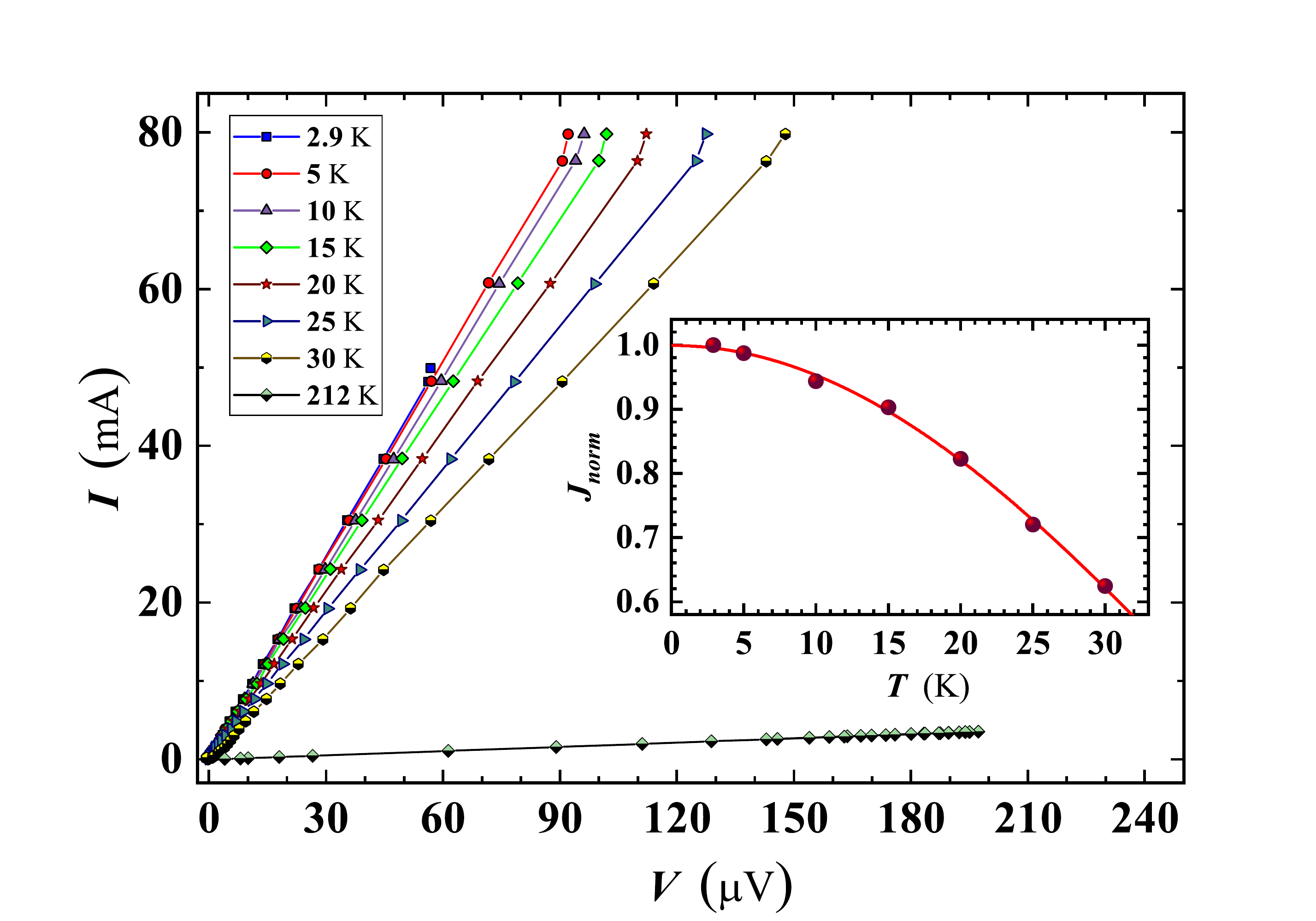}
\caption{{Current-voltage characteristics of a doped KPT pellet at several fixed temperatures. Inset: temperature dependence of the normalised current density defined as $J(T)/J(T_0)$ where $J(T)$ and~$J(T_0)$ are the current densities values, for a measured voltage drop of 90 $\mu$V, at $T$ and $T_0 = 2.9$ K, respectively.  The red line is the least squared fit of the experimental points by Equation (\ref{eqJc}) (see the text) resulting in a $T_C \simeq 73$~K.} }
\label{fig_I-V_T}
\end{figure}

The nature of the detected fall of $\rho(T)$ was investigated by carrying out I-V characteristics at several fixed temperature. 
At $T \leq 30$ K, I-V characteristics exhibits a linear behaviour even if the voltage drop remains relatively low (in the range of tens of microvolts) up to about 80 mA, close~to the maximum intensity achievable by the current source used in our system of measure.
We~found that the temperature dependence of the current density value at 90 $\mu$V
, exhibits a smooth continuous decrease, raising $T$. We have carried out a least squared fit by a Ginzburg--Landau-like function, considering as dependent variable the normalised current density, $J_{norm}$, here defined as the ratio between the current density at $T$, $J(T)$, and the current density at the lowest temperature  (i.e., $T_0 = 2.9$ K), $J(T_0)$:

\begin{equation}
J_{norm}(T) = \left[1-(T/T_C)^2\right]^{3/2}\left[1+(T/T_C)^2\right]^{1/2}
\label{eqJc}
\end{equation}

Application of Equation (\ref{eqJc}) has allowed estimating a superconducting critical temperature of~$T_C \simeq 73$ K. This interesting result and the good best fitting by Equation (\ref{eqJc}) of the experimental points $J_{norm}(T)$ (Figure \ref{fig_I-V_T}, inset) could suggest a proportionality between measured current density and the superconducting critical current density of the KPT sample. 

Our finding of a bell shaped curve of the KPT resistivity suggests two different temperature dependent conduction mechanisms. Below RT, down to $\approx$ 150 K, 
$\rho(T)$ presents a dielectric-like character, with an activation energy progressively decreasing from 1.29 eV, around room temperature, to about 2 meV, at $\simeq$ 150 K (Figure \ref{fig_rho_2}, inset). 
This behaviour could be explained assuming a~trapping mechanism of charge carriers, as reported for polycrystalline thin films of $p$-terphenyl~\cite{Lipinski}. 
Moreover,~the low value of $\rho(T)$ at RT could suggest a partial contribution of $K^+$ ions to the electrical conduction that, anyway, is expected to rapidly decrease lowering $T$.
Below $\approx$ 130 K, $\rho(T)$ becomes metallic-like indicating an enhanced contribution to the conduction by electrons. The initial slow decrease of $\rho(T)$ is  followed by a steeper fall at $\approx$ 90 K that, in addition to the residual value of {$10^{-2}\div10^{-3}$ $\Omega$cm below 30 K} (Figure \ref{fig_rho_2}), {looks like that of a superconducting transition, that depending on the criterium taken into account,} a superconducting transition temperature of $T_C \simeq 94$ K \cite{Pinto2018} or $T_C \simeq 87$ K (i.e. the $T$ corresponding to the 50\% of the flat region of the $\rho(T)$ curve, here equals to 79~$\Omega$cm) can be defined, with a transition width of $\Delta T_C \simeq 22$ K, in both cases. 
The region between 150~K and 130 K marks the transition between these two conduction regimes.

The non-monotonic behaviour of the KPT resistivity shares important similarities with that observed in K$_x$-picene, an organic molecule formed by 5 benzene rings fused together along their edges and doped by K-ions ($x\simeq 3$) \cite{Teranishi2013}. Picene and $p$-terphenyl molecules, can be considered members of the same family of hydrocarbon superconductors.

Several findings measured in K$_x$-picene pellets deserves to be mentioned and compared to those of KPT: a similar behaviour of the $\rho(T)$ curve, with an initial increase of $\rho(T)$ followed by a rapid decrease, lowering $T$ from RT; the highest resistivity value at the curve peak is 100 $\Omega$cm ($\simeq 80$ $\Omega$cm in KPT); $\rho_{RT} \simeq 0.4$  $\Omega$cm (a few $\Omega$cm in KPT); a superconducting transition width several kelvin wide ($\simeq 22$ K in KPT).

The transition width, in particular, and the {huge drop} of the KPT resistivity, {are assigned to} the granular nature of the KPT disks, fabricated by a compressed powder as for the K$_x$-picene pellets~\cite{Teranishi2013}. 
{Anyway, possible evidence of superconductivity in KPT compounds was inferred by magnetic susceptibility only \cite{Chen3,Neha2018}, while ARPES and STM have experimentally proved the existence of a~large superconducting energy gap \cite{Dessau,Ren}, while previous resistivity measurements gave negative results~\cite{Carrera2019}. On the other hand, magnetic susceptibility measurements indicating a superconducting state were reported for all hydrocarbon superconductors, including K$_x$-picene \cite{Kubozono2015}. The resistivity drop by several orders of magnitude here observed in KPT represents a first possible direct evidence for superconductivity in this compound. }

To successfully measure electrical properties of KPT, a percolating path connecting the~electrical contacts is necessary. Hence, a large number of grains in the superconducting state will be involved to detect a drop of the resistivity. On the contrary, in the magnetic susceptibility, the~superconducting features can be produced by not interconnected regions of the material giving more easily evidence of supercondutivity \cite{Chen3,Neha2018}.
Our results support the hypothesis of a~superconductivity {state} confined in isolated regions of the KPT sample
that have probably prevented till now to observe any superconducting transition by electrical measurements.
This is also due to the quite small values of the coherence length of the order of tens of nanometers that, \mbox{due to the Josephson} tunneling of Cooper pairs, would weakly couple KPT grains with a size of the order of microns.
The~possibility to observe a superconducting transition in the whole KPT disk, by electronic transport measurements,
will depend on several factors such as purity of the pristine material, dimension of KPT crystallites, their density and distribution in the dielectric matrix. Achieving high-$T_C$ superconductivity will depend mainly on the doping level of the $p$-terphenyl molecule.
Our Raman analysis has confirmed both the high quality and pureness of the undoped $p$-terphenyl and a successfull doping of KPT as well, that the AAS technique has established, in a quantitative way, \mbox{to be of} $\simeq 2$ K-ions for each $p$-terphenyl molecule.
Additionally, the pressure applied to a KPT disk is expected to affect its electrical properties, due to an improved contact between KPT grains.

To explain the electrical behaviour detected below $\approx$ 130 K, we hypothesise that KPT disks could be made of superconducting crystallites, with a distribution in their taille and $T_C$, dispersed \mbox{in a dielectric} matrix, as suggested by the presence of low-intensity Raman peaks due to residual
grains of the undoped phase (see Figure \ref{figRm4}, $T=203$ K). At low $T$ ($< 130$ K), under sufficient mechanical pressure, superconducting grains could have formed a percolative path between the metal contacts. Due to the $T_C$ distribution, the system will not undergo a narrow superconducting transition but will progressively reduce the resistivity, upon superconducting transition of single KPT crystallites.
\mbox{It is reasonable} to suppose that at the lowest temperature not all crystallites have become superconducting or local inhomogeneity (see the spectrum at 203 K in Figure \ref{figRm4}) has prevented a~drop to zero of~$\rho(T)$ \cite{Teranishi2013} {even if the lowest attained $\rho(T)$ values for our KPT material are \mbox{$10^{-2}\div10^{-3}$ $\Omega$cm}}. This last hypothesis is further supported by the lacking of a sudden transition to the normal state in the I-V curves, here substituted by a {linear} change of $I$ even at the lowest $T$ (see Figure \ref{fig_I-V_T}).
Both the shape of the transition and the lacking of a negligible $\rho(T)$ at $T_C$ can be considered a manifestation of the granular nature of the KPT material \cite{Teranishi2013}. {It is worthwhile noting that $T_C$ derived from I-V curves, fitting $J_{norm}$ by Equation (\ref{eqJc}), results comparable to that derived from resistivity measurements.}

Percolative superconductivity is typical in granular superconducting materials \cite{Ponta2013}.
Please note that intense Raman peaks, detected in the high frequency region of the spectrum of the KPT powder, even at $T$ = 108 K, justify a possible key role played by high energy phonons in the superconducting pairing.
Even though the mechanism driving Cooper pairing in KPT systems is still unknown, the electron-phonon coupling with high energy phonons should play a key role in the high temperature superconductivity, as discussed above. In this framework, the density of electronic states (DOS) close to the Fermi energy is expected to influence the value of the critical temperature. As shown in {Figure 2} of Ref. \cite{Geilhufe2018} the DOS of $p$-terphenyl has several peaks associated with van Hove singularities, because of multiple bands characterizing the anisotropic electronic structure. Electron doping by K atoms, increasing the Fermi energy toward positive values (right part of the DOS in {Figure 2}
 of Ref. \cite{Geilhufe2018}), will determine a crossing of the van Hove singularities and an increase of the superconducting critical temperature. 
Our doping level of K-atoms, intermediate between the one in Ref.~\cite{Wang2017} ($T_c = 7.2$ K) and the other in Ref.~\cite{Chen3} ($T_c = 123$ K) is compatible with the intermediate value of $T_C$ around 90 K found in this work.

\section{Conclusions}
To conclude, the huge suppression of the resistivity of K-doped $p$-terphenyl measured in our experiments below $T$ = 90 K, {of about four orders} of magnitude in a temperature range of $\simeq 20$ K, {represents a possible} signature of a superconducting transition in this metallo-organic compound, providing additional support to the evidence of high-$T_C$ superconductivity found by magnetic susceptibility, ARPES and STM measurements. Future experiments on the electrical transport properties of KPT materials will focus on the magnetic field dependence of the resistivity curve $\rho(T,H)$, to study the expected suppression of the critical temperature when the external magnetic field is increased toward its upper critical value. Still we have to underline that this is a complex quantum material and samples are characterized by inhomogeneous concentrations of dopants and by a mixture of different coexisting phases.

\section{Acknowledgements}
We thank David Neilson, Javid Rezvani and Meenakshi Sharma for useful discussions. We acknowledge the University of Camerino for providing technical and financial support. The School of Science and Technology and the Physics Division are acknowledged for their contribution to the installation of the microRaman and the Helium closed cycle cryostat equipments.


\begin{thebibliography}{0}
\expandafter\ifx\csname natexlab\endcsname\relax\def\natexlab#1{#1}\fi
\expandafter\ifx\csname bibnamefont\endcsname\relax
  \def\bibnamefont#1{#1}\fi
\expandafter\ifx\csname bibfnamefont\endcsname\relax
  \def\bibfnamefont#1{#1}\fi
\expandafter\ifx\csname citenamefont\endcsname\relax
  \def\citenamefont#1{#1}\fi
\expandafter\ifx\csname url\endcsname\relax
  \def\url#1{\texttt{#1}}\fi
\expandafter\ifx\csname urlprefix\endcsname\relax\def\urlprefix{URL }\fi
\providecommand{\bibinfo}[2]{#2}
\providecommand{\eprint}[2][]{\url{#2}}

\end{thebibliography}


\begin{thebibliography}{-------}
\providecommand{\natexlab}[1]{#1}
\bibitem[Romero \em{et~al.}(2017)Romero, Pitcher, Hiley, Whitehead, Kar, Ganin,
  Antypov, Collins, Dyer, Klupp, Colman, Prassides, and Rosseinsky]{Romero2017}
Romero, F.D.; Pitcher, M.J.; Hiley, C.I.; Whitehead, G.F.S.; Kar, S.; Ganin,
  A.Y.; Antypov, D.; Collins,~C.; Dyer,~M.S.; Klupp, G.; Colman, R.H.;
  Prassides, K.; Rosseinsky, M.J. Redox-controlled potassium intercalation into two polyaromatic
  hydrocarbon solids. {\em Nat. Chem.} {\bf 2017}, {\em 9},~644. doi:10.1038/nchem.2765.

\bibitem[R.~Mitsuhashi \em{et~al.}(2010)R.~Mitsuhashi, Yamanari, Mitamura,
  Kambe, Ikeda, Okamoto, Fujiwara, Yamaji, Kawasaki, Maniwa, and
  Kubozono]{Mitsuhashi2010}
Mitsuhashi, R; Suzuki, Y.; Yamanari, Y.; Mitamura, H.; Kambe, T.; Ikeda, N.; Okamoto,
  H.; Fujiwara, A.; Yamaji,~M.; Kawasaki, N.; Maniwa, Y.; Kubozono, Y.
 Superconductivity in alkali-metal-doped picene.
{\em Nature} {\bf 2010}, {\em 464},~76--79.

\bibitem[M. \em{et~al.}(1970)M., Rietveld, Maslen, and Clews]{Rietveld}
Rietveld, H.M.; Maslen, E.N.; Clews, C.J.B.
An x-ray and neutron diffraction refinement of the structure of
  $p$-terphenyl.
{\em Acta Crystal. B Struct. Crystallog. Cryst. Chem.} {\bf 1970},
  {\em 26},~693. doi:10.1107/S0567740870003023.

\bibitem[Baudour \em{et~al.}(1977)Baudour, Cailleau, and Yelon]{Baudour}
Baudour, J.L.; Cailleau, H.; Yelon, W.B.
Structural phase transition in polyphenyls. IV. double-well potential
  in the disordered phase of $p$-terphenyl from neutron (200 K) and x-ray
  (room-temperature) diffraction data.
{\em Acta Crystal. B Struct. Crystallog. Cryst. Chem.} {\bf 1977},
  {\em 33},~1773. doi:10.1107/S0567740877007043.

\bibitem[Cailleau \em{et~al.}(1979)Cailleau, Heidemann, and Zeyen]{Cailleau1}
Cailleau, H.; Heidemann, A.; Zeyen, C.M.E.
Observation of critical slowing down at the structural phase
  transition in $p$-terphenyl by high-resolution neutron spectroscopy.
{\em J. Phys. C Solid State Phys.} {\bf 1979}, {\em 12},~L411. doi:10.1088/0022-3719/12/11/002.

\bibitem[Lechner \em{et~al.}(1984)Lechner, Toudic, and Cailleau]{Lechner}
Lechner, R.E.; Toudic, B.; Cailleau, H.
Observation of the effects of critical phenomena in paraterphenyl on
  quasielastic incoherent neutron spectra.
{\em J. Phys. C Sol. State Phys.} {\bf 1984}, {\em 17},~405. doi:10.1088/0022-3719/17/3/01.

\bibitem[Cailleau \em{et~al.}(1980)Cailleau, Baudour, Meinnel, Dworkin, Moussa,
  and Zeyen]{Cailleau2}
Cailleau, H.; Baudour, J.L.; Meinnel, J.; Dworkin, A.; Moussa, F.; Zeyen,
  C.M.E.
 Double-well potentials and structural phase-transitions in
  polyphenyls.
 {\em Faraday Discus. Chem. Soc.} {\bf 1980}, {\em 69},~7. doi:10.1039/dc9806900007.

\bibitem[Baranyai and Welberry(1992)]{Baranyai}
Baranyai, A.; Welberry, T.R.
 Molecular dynamics simulation study of solid polyphenyls: Structures
  determined by the interplay between intra- and intermolecular forces.
 {\em Mol. Phys.} {\bf 1992}, {\em 75},~867. doi:10.1080/00268979200100661.

\bibitem[Goossens \em{et~al.}(2009)Goossens, Beasley, Welberry, Gutmann, and
  Piltz]{Goossens}
Goossens, D.J.; Beasley, A.G.; Welberry, T.R.; Gutmann, M.J.; Piltz, R.O.
 Neutron diffuse scattering in deuterated para-Terphenyl, C 18 D 14.
 {\em J. Phys. Cond. Mat.} {\bf 2009}, {\em 21},~124204.

\bibitem[Rice \em{et~al.}(2013)Rice, Tham, and Chronister]{Rice}
Rice, A.P.; Tham, F.S.; Chronister, E.L.
 A temperature dependent x-ray study of the Order–Disorder
  enantiotropic phase transition of $p$-terphenyl.
 {\em J. Chem. Crystal.} {\bf 2013}, {\em 43},~14. doi:10.1007/s10870-012-0378-6.

\bibitem[Pinto \em{et~al.}(2016)Pinto, Rezvani, Favre, Berbezier, Fretto, and
  Boarino]{Pinto2016}
Pinto, N.; Rezvani, S.J.; Favre, L.; Berbezier, I.; Fretto, M.; Boarino, L.
 {Geometrically induced electron-electron interaction in semiconductor
  nanowires}.
 {\em Appl. Phys. Lett.} {\bf 2016}.
doi:10.1063/1.4962893.

\bibitem[Rezvani \em{et~al.}(2016)Rezvani, Gunnella, Neilson, Boarino, Croin,
  Aprile, Fretto, Rizzi, Antonioli, and Pinto]{Rezvani2016}
Rezvani, S.J.; Gunnella, R.; Neilson, D.; Boarino, L.; Croin, L.; Aprile, G.;
  Fretto, M.; Rizzi, P.; Antonioli, D.; Pinto, N.
 {Effect of carrier tunneling on the structure of Si nanowires
  fabricated by metal assisted etching}.
{\em Nanotechnology} {\bf 2016}.
  doi:10.1088/0957-4484/27/34/345301.

\bibitem[Mazziotti \em{et~al.}(2017)Mazziotti, Valletta, Campi, Innocenti,
  Perali, and Bianconi]{Mazziotti}
Mazziotti, M.V.; Valletta, A.; Campi, G.; Innocenti, D.; Perali, A.; Bianconi,
  A. Possible Fano resonance for high-Tc multi-gap superconductivity in
  p-Terphenyl doped by K at the Lifshitz transition.
 {\em {EPL} (Europhys. Lett.)} {\bf 2017}, {\em 118},~37003.
  doi:10.1209/0295-5075/118/37003.

\bibitem[Barba \em{et~al.}(2018)Barba, Chita, Campi, Suber, Bauer, Marcelli,
  and Bianconi]{Bianconi1}
Barba, L.; Chita, G.; Campi, G.; Suber, L.; Bauer, E.M.; Marcelli, A.;
  Bianconi, A.
Anisotropic thermal expansion of $p$-terphenyl: A self-assembled
  supramolecular array of poly-p-phenyl nanoribbons.
{\em J.~Supercond. Novel Magnet.} {\bf 2018}, {\em 31},~703. doi:10.1007/s10948-017-4407-8.

\bibitem[Wang \em{et~al.}()Wang, Gao, Huang, , and Chen]{Wang2017}
Wang, R.S.; Gao, Y.; Huang, Z.B.; Chen, X.J.
 Superconductivity in $p$-terphenyl. \emph{arXiv} \textbf{2017}, arXiv:1703.05803.

\bibitem[Huang \em{et~al.}(2019)Huang, Zhong, Wang, Han, Lin, and
  Chen]{Huang2019}
Huang, G.; Zhong, G.H.; Wang, R.S.; Han, J.X.; Lin, H.Q.; Chen, X.J.
 Superconductivity and phase stability of potassium-doped
  p-quinquephenyl.
 {\em Carbon} {\bf 2019}, {\em 143},~837. doi:10.1016/j.carbon.2018.12.001.

\bibitem[Yan \em{et~al.}(2019)Yan, Zhong, Wang, Zhang, Lin, and Chen]{Yan2019}
Yan, J.F.; Zhong, G.H.; Wang, R.S.; Zhang, K.; Lin, H.Q.; Chen, X.J.
Superconductivity and Phase Stability of Potassium-Intercalated
  p-Quaterphenyl.
 {\em J. Phys. Chem. Lett.} {\bf 2019}, {\em 10},~40. doi:10.1021/acs.jpclett.8b03263.

\bibitem[Zhong \em{et~al.}(2018)Zhong, Yang, Zhang, Wang, Zhang, Lin, , and
  Chen]{Zhong2018a}
Zhong, G.H.; Yang, D.Y.; Zhang, K.; Wang, R.S.; Zhang, C.; Lin, H.Q.;  Chen,
  X.J.
 Superconductivity and phase stability of potassium-doped biphenyl.
{\em Phys. Chem. Chem. Phys.} {\bf 2018}, {\em 20},~25217. doi:10.1039/C8CP05184D.

\bibitem[Gao \em{et~al.}(2016)Gao, Wang, Wu, Cheng, Deng, X.-W.Yan, and
  Huang]{Chen1}
Gao, Y.; Wang, R.S.; Wu, X.L.; Cheng, J.; Deng, T.G.; Yan, X.-W.; Huang, Z.B.
Searching superconductivity in potassium-doped $p$-terphenyl.
{\em Acta Phys. Sin.} {\bf 2016}, {\em 65},~077402.

\bibitem[Wang \em{et~al.}({\natexlab{a}})Wang, Gao, Huang, and Chen]{Chen2}
Wang, R.S.; Gao, Y.; Huang, Z.B.; Chen, X.J.
 Superconductivity at 43 K in a single C-C bond linked terphenyl. \emph{arXiv} \textbf{2017},
 arXiv:1703.05804.

\bibitem[Wang \em{et~al.}({\natexlab{b}})Wang, Gao, Huang, and Chen]{Chen3}
Wang, R.S.; Gao, Y.; Huang, Z.B.; Chen, X.J.
 Superconductivity above 120 kelvin in a chain link molecule. \emph{arXiv} \textbf{2017},
arXiv: 1703.06641.

\bibitem[Liu \em{et~al.}(2017)Liu, Lin, Kang, Zhang, Zhu, and Wen]{Liu1}
Liu, W.; Lin, H.; Kang, R.; Zhang, Y.; Zhu, X.; Wen, H.H.
 Magnetization of potassium doped $p$-terphenyl and p-quaterphenyl by
  high pressure synthesis.
 {\em Phys. Rev. B} {\bf 2017}, {\em 96},~224501. doi:10.1103/PhysRevB.96.224501.

\bibitem[Neha \em{et~al.}(2018)Neha, Bhardwaj, Sahu, and Patnaik]{Neha2018}
Neha, P.; Bhardwaj, A.; Sahu, V.; Patnaik, S.
 Facile synthesis of potassium intercalated $p$-terphenyl and
  signatures of a possible high T$_c$ phase.
 {\em Phys. C Supercond. App.} {\bf 2018}, {\em 554},~1. doi:10.1016/j.physc.2018.08.011.

\bibitem[Carrera \em{et~al.}(2019)Carrera, McDonald, Untiedt,
  Garc{\'\i}a-Hern{\'a}ndez, Mompean, A.~Verg{\'e}s, and Guijarro]{Carrera2019}
Carrera, M.; McDonald, J.L.; Untiedt, C.; Garc{\'\i}a-Hern{\'a}ndez, M.;
  Mompean, F.; Verg{\'e}s, J.A.; Guijarro,~A.
Characterization of Main Phase in K$_x$p-Terphenyl and Its Largest
  Congener K$_x$poly(p-phenylene): A Report of Their Magnetic and Electric
  Properties.
 {\em J. Phys. Chem. C} {\bf 2019}, {\em 123},~5264--5272.
  doi:10.1021/acs.jpcc.8b12475.

\bibitem[Li \em{et~al.}(2019)Li, Zhou, Parham, Nummy, Griffith, Gordon,
  Chronister, and Dessau]{Dessau}
Li, H.; Zhou, X.; Parham, S.; Nummy, T.; Griffith, J.; Gordon, K.; Chronister,
  E.L.; Dessau, D.S.
 Spectroscopic evidence of low-ernergy gaps persisting up to 120
  Kelvin in surface-doped $p$-terphenyl cristals.
 \mbox{{\em Phys. Rev.~B}} {\bf 2019}, {\em 100},~064511. doi:10.1103/PhysRevB.100.064511.

\bibitem[Ren \em{et~al.}(2019)Ren, Chen, Liu, Chen, Qiao, Chen, Zhou, Zhang,
  Yan, and Feng]{Ren}
Ren, M.Q.; Chen, W.; Liu, Q.; Chen, C.; Qiao, Y.J.; Chen, Y.J.; Zhou, G.;
  Zhang, T.; Yan, Y.J.; Feng, D.L.
Observation of novel gapped phases in potassium doped single layer
  $p$-terphenyl on Au (111).
\mbox{{\em Phys. Rev. B}} {\bf 2019}, {\em 99},~045417. doi:10.1103/PhysRevB.99.045417.

\bibitem[Perali \em{et~al.}(2002)Perali, Pieri, Strinati, and
  Castellani]{Perali1}
Perali, A.; Pieri, P.; Strinati, G.C.; Castellani, C.
 Pseudogap and spectral function from superconducting fluctuations to
  the bosonic limit.
 {\em Phys. Rev. B} {\bf 2002}, {\em 66},~024510. doi:10.1103/PhysRevB.66.024510.

\bibitem[Palestini \em{et~al.}(2012)Palestini, Perali, Pieri, and
  Strinati]{Palestini2012}
Palestini, F.; Perali, A.; Pieri, P.; Strinati, G.C.
 Dispersions, weights, and widths of the single-particle spectral
  function in the normal phase of a Fermi gas.
 {\em Phys. Rev. B} {\bf 2012}, {\em 85},~024517. doi:10.1103/PhysRevB.85.024517.

\bibitem[Gaebler \em{et~al.}(2010)Gaebler, Stewart, Drake, Jin, Perali, Pieri,
  and Strinati]{Perali2}
Gaebler, J.P.; Stewart, J.T.; Drake, T.E.; Jin, D.S.; Perali, A.; Pieri, P.;
  Strinati, G.C.
 Observation of pseudogap behaviour in a strongly interacting Fermi
  gas.
 {\em Nat. Phys.} {\bf 2010}, {\em 6},~569. doi:10.1038/nphys1709.

\bibitem[Marsiglio \em{et~al.}(2015)Marsiglio, Pieri, Perali, Palestini, and
  Strinati]{Marsiglio2015}
Marsiglio, F.; Pieri, P.; Perali, A.; Palestini, F.; Strinati, G.C.
 Pairing effects in the normal phase of a~two-dimensional Fermi gas.
 {\em Phys. Rev. B} {\bf 2015}, {\em 91},~054509. doi:10.1103/PhysRevB.91.054509.

\bibitem[Salasnich \em{et~al.}(2019)Salasnich, Shanenko, Vagov, Aguiar, and
  Perali]{Salasnich2019}
Salasnich, L.; Shanenko, A.A.; Vagov, A.; Aguiar, J.A.; Perali, A.
 Screening of pair fluctuations in superconductors with coupled
  shallow and deep bands: A route to higher-temperature superconductivity.
 {\em Phys. Rev. B} {\bf 2019}, {\em 100},~064510. doi:10.1103/PhysRevB.100.064510.

\bibitem[Tajima \em{et~al.}(2019)Tajima, Yerin, Perali, and Pieri]{Tajima2019}
Tajima, H.; Yerin, Y.; Perali, A.; Pieri, P.
 Enhanced critical temperature, pairing fluctuation effects, and
  BCS-BEC crossover in a two-band Fermi gas.
 {\em Phys. Rev. B} {\bf 2019}, {\em 99},~180503(R). doi:10.1103/PhysRevB.99.180503.

\bibitem[Valletta \em{et~al.}(1997)Valletta, Bianconi, Perali, and
  Saini]{Valletta1997}
Valletta, A.; Bianconi, A.; Perali, A.; Saini, N.L.
 Electronic and superconducting properties of a superlattice of
  quantum stripes at the atomic limit.
 {\em Z. Phys. B} {\bf 1997}, {\em 104},~707. doi:10.1007/s002570050513.

\bibitem[Fabrizio \em{et~al.}()Fabrizio, Qin, Naghavi, and
  Tosatti]{Fabrizio2017}
Fabrizio, M.; Qin, T.; Naghavi, S.S.; Tosatti, E.
 Two-Band s$\pm$ Strongly Correlated Superconductivity in K3
  $p$-Terphenyl ? \emph{arXiv} \textbf{2017}, arXiv:1705.05066.

\bibitem[Connelly and Geiger(1996)]{Connelly1996}
Connelly, N.G.; Geiger, W.E.
 Chemical redox agent for organometallic chemistry.
 {\em Chem. Rev.} {\bf 1996}, {\em 96},~877. doi:10.1021/cr940053x.

\bibitem[Zhong \em{et~al.}(2018)Zhong, Wang, Wang, Han, Zhang, Chen, and
  Lin]{Zhong2018}
Zhong, G.H.; Wang, X.H.; Wang, R.S.; Han, J.X.; Zhang, C.; Chen, X.J.; Lin,
  H.Q.
 Structural and Bonding Characteristics of Potassium-Doped
  $p$-Terphenyl Superconductors.
 {\em J. Phys. Chem. C} {\bf 2018}, {\em 122},~3801. doi:10.1021/acs.jpcc.7b12616.

\bibitem[Pinto \em{et~al.}(2018)Pinto, Rezvani, Perali, Flammia, Milosevic,
  Fretto, Cassiago, and Leo]{Pinto2018}
Pinto, N.; Rezvani, S.J.; Perali, A.; Flammia, L.; Milosevic, M.V.; Fretto, M.;
  Cassiago, C.; Leo, N.D.
 Dimensional crossover and incipient quantum size effects in
  superconducting niobium nanofilms.
 {\em Sci. Rep.} {\bf 2018}, {\em 8},~4710. doi:10.1038/s41598-018-22983-6.

\bibitem[Rezvani \em{et~al.}(2019)Rezvani, Perali, Fretto, Leo, Flammia,
  Milosevic, Nannarone, and Pinto]{Rezvani2019}
Rezvani, S.J.; Perali, A.; Fretto, M.; Leo, N.D.; Flammia, L.; Milosevic, M.V.;
  Nannarone, S.; Pinto, N.
 Substrate-Induced Proximity Effect in Superconducting Niobium
  Nanofilms.
 {\em Cond. Matter} {\bf 2019}, {\em 4},~4. doi:10.3390/condmat4010004.

\bibitem[Bolton and Prasad(1978)]{Bolton1978}
Bolton, B.A.; Prasad, P.N.
 Phase transitions in polyphenyls: Raman spectra of $p$-terphenyl and
  p-quaterphenyl in the solid state.
 {\em Chem. Phys.} {\bf 1978}, {\em 35},~331. doi:10.1016/S0301-0104(78)85248-3.

\bibitem[da~Costa \em{et~al.}(1997)da~Costa, Amado, Becucci, and
  Kryschi]{Amorim_da_Costa1997}
da~Costa, A.A.; Amado, A.; Becucci, M.; Kryschi, C.
 Order-disorder phase transition in $p$-terphenyl and $p$-terphenyl:
  Tetracene doped crystals as studied by Raman spectroscopy.
 {\em J. Mol. Struct.} {\bf 1997}, {\em 416},~69. doi:10.1016/S0022-2860(97)00048-3.

\bibitem[Zhang \em{et~al.}(2018)Zhang, Wang, and Chen]{Zhang2018}
Zhang, K.; Wang, R.; Chen, X.
 Vibrational Properties of $p$-terphenyl.
 {\em J. Phys. Chem. A} {\bf 2018}, {\em 122},~6903. doi:10.1021/acs.jpca.8b05462.

\bibitem[Baudour \em{et~al.}(1976)Baudour, Delugeard, and
  Cailleau]{Baudour1976}
Baudour, J.L.; Delugeard, Y.; Cailleau, H.
 {Transition structurale dans les polyph{\'{e}}nyles. I. Structure
  cristalline de la phase basse temp{\'{e}}rature du {\it p}-terph{\'{e}}nyle
  {\`{a}} 113 K}.
 {\em Acta Crystallogr. Sect. B} {\bf 1976}, {\em
  32},~150--154.
  doi:10.1107/S0567740876002501.

\bibitem[Brédas \em{et~al.}(1984)Brédas, Thémans, Fripiat, André, and
  Chance]{Bredas1984}
Brédas, J.; Thémans, B.; Fripiat, J.; André, J.; Chance, J.R.
 An ab-initio study of the geometry and electronic-structure
  modifications upon doping.
 {\em Phys. Rev. B} {\bf 1984}, {\em 29},~6761. doi:10.1103/PhysRevB.29.6761.

\bibitem[Furukawa(1996)]{Furukawa1996}
Furukawa, Y.
 Electronic Absorption and Vibrational Spectroscopies of Conjugated
  Conducting Polymers.
{\em J.~Phys. Chem.} {\bf 1996}, {\em 100},~15644. doi:10.1021/jp960608n.

\bibitem[Péres \em{et~al.}(2005)Péres, Spiesser, and Froyer]{Peres2005}
Péres, L.; Spiesser, M.; Froyer, G.
 Reduction of $p$-terphenyl, p-quaterphenyl and p-sexiphenyl using
  alkali metal in liquid ammonia: Process and characterization of the reduced
  compounds.
 {\em Synth. Met.} {\bf 2005}, {\em 155},~450. doi:10.1016/j.synthmet.2005.02.020.

\bibitem[Furukawa \em{et~al.}(1993)Furukawa, Ohtsuka, and Tasumi]{Furukawa1993}
Furukawa, Y.; Ohtsuka, H.; Tasumi, M.
 Raman studies of polarons and bipolarons in sodium-doped
  poly-p-phenylene.
 {\em Synth. Met.} {\bf 1993}, {\em 55},~516. doi:10.1016/0379-6779(93)90984-5.

\bibitem[Dubois \em{et~al.}(2003)Dubois, Froyer, Louarn, and
  Billaud]{Dubois2003}
Dubois, M.; Froyer, G.; Louarn, G.; Billaud, D.
 Raman spectroelectrochemical study of sodium intercalation into
  poly(p-phenylene).
{\em Spectrochim. Acta A Mol. Biomol. Spectros.} {\bf 2003},
  {\em 59},~1849. doi:10.1016/S1386-1425(02)00417-1.

\bibitem[Chakraverty(1981)]{Chak1981}
Chakraverty, B.K.
 Bipolarons and superconductivity.
{\em J. Phys.} {\bf 1981}, {\em 42},~1351. doi:10.1051/jphys:019810042090135100.

\bibitem[Alexandrov and Ranninger(1981)]{Alexandrov1981}
Alexandrov, A.; Ranninger, J.
 Bipolaronic superconductivity.
 {\em Phys. Rev. B} {\bf 1981}, {\em 24},~1164. doi:10.1103/PhysRevB.24.1164.

\bibitem[Lipinski \em{et~al.}(1980)Lipinski, Mycielski, and Swiatek]{Lipinski}
Lipinski, A.; Mycielski, W.; Swiatek, J.
 Charge carrier transport and d.c. conductivity in thin
  polycrystalline $p$-terphenyl films.
 {\em J. Phys. Chem. Solids} {\bf 1980}, {\em 41},~455. doi:10.1016/0022-3697(80)90174-2.

\bibitem[Staryga and Swiatek(1979)]{Staryga}
Staryga, E.; Swiatek, J.
 The electrical conductivity in thin polycrystalline $p$-terphenyl
  films.
 {\em Thin Sol. Films} {\bf 1979}, {\em 56},~311. doi:10.1016/0040-6090(79)90132-9.

\bibitem[Tkaczyk(1999)]{Tkaczyk}
Tkaczyk, S.W.
Electrical conductivity of the polycrystalline films of
  $p$-terphenyl.
 {\em Proc. SPIE} {\bf 1999}, {\em 3725},~232. doi:10.1117/12.344742.

\bibitem[Bizzarri \em{et~al.}(1973)Bizzarri, Casa, and Pietra]{Bizzarri}
Bizzarri, P.C.; Casa, C.D.; Pietra, S.
 Electrical conductivity of o-, m-, and $p$-terphenyls.
 {\em Z. Naturforsch} {\bf 1973},~{\em 28b},~331. doi:10.1515/znb-1973-5-621.

\bibitem[Teranishi \em{et~al.}(2013)Teranishi, He, Sakai, Izumi, Goto, Eguchi,
  Takabayashi, Kambe, and Kubozono]{Teranishi2013}
Teranishi, K.; He, X.; Sakai, Y.; Izumi, M.; Goto, H.; Eguchi, R.; Takabayashi,
  Y.; Kambe, T.; Kubozono, Y.
 Observation of zero resistivity in K-doped picene.
 {\em Phys. Rev. B} {\bf 2013}, {\em 87},~060505. doi:10.1103/PhysRevB.87.060505.

\bibitem[Kubozono \em{et~al.}(2015)Kubozono, Goto, Jabuchi, Yokoya, Kambe,
  Sakai, Izumi, Zheng, Hamao, Nguyen, Sakata, Kagayama, and
  Shimizu]{Kubozono2015}
Kubozono, Y.; Goto, H.; Jabuchi, T.; Yokoya, T.; Kambe, T.; Sakai, Y.; Izumi,
  M.; Zheng, L.; Hamao, S.; Nguyen, H.L.; Sakata, M.; Kagayama, T.; Shimizu, K.
 Superconductivity in aromatic hydrocarbons.
 {\em Physica~C} {\bf 2015}, {\em 514},~199. doi:10.1016/j.physc.2015.02.015.

\bibitem[Ponta \em{et~al.}(2013)Ponta, Andreoli, and Carbone]{Ponta2013}
Ponta, L.; Andreoli, V.; Carbone, A.
 Superconducting-insulator transition in disordered Josephson
  junctions networks.
 {\em Eur. Phys. J. B} {\bf 2013}, {\em 86},~24. doi:10.1140/epjb/e2012-30216-x.

\bibitem[Geilhufe \em{et~al.}(2018)Geilhufe, Borysov, Kalpakchi, and
  Balatsky]{Geilhufe2018}
Geilhufe, R.M.; Borysov, S.S.; Kalpakchi, D.; Balatsky, A.V.
 Towards novel organic high-Tc superconductors: Data mining using
  density of states similarity search.
 {\em Phys. Rev. Mater.} {\bf 2018}, {\em 2},~024802. doi:10.1103/PhysRevMaterials.2.024802.

\end{thebibliography}
\end{document}